\documentclass[final,authoryear,5p,times,twocolumn]{elsarticle}

\usepackage{amssymb}

\journal{Icarus}
\bibpunct{(}{)}{;}{a}{}{,}
\usepackage{graphicx,subfigure}

\usepackage{ulem}
\usepackage{pstricks,hyperref}

\begin{document}

\begin{frontmatter}

\title{Mars Encounters cause fresh surfaces on some near-Earth asteroids}

\author[mit]{Francesca E. DeMeo}
\author[mit]{Richard P. Binzel}
\author[mit,swe]{Matthew Lockhart}

\address[mit]{Department of Earth, Atmospheric, and Planetary Sciences, Massachusetts Institute of Technology, 77 Massachusetts Avenue, Cambridge, MA 02139 USA}
\address[swe]{Department of Physics and Astronomy, Uppsala University, Box 516, 751 20 Uppsala, Sweden}

%%%%%%%%%%%%%%%%%%%%%%%%%%%%%%%%

\begin{abstract}
All airless bodies are subject to the space environment, and spectral differences between asteroids and meteorites suggest many asteroids become weathered on very short ($<$1My) timescales. The spectra of some asteroids, particularly Q-types, indicate surfaces that appear young and fresh, implying they have been recently been exposed. Previous work found that Earth encounters were the dominant freshening mechanism and could be responsible for all near-Earth object (NEO) Q-types. In this work we increase the known NEO Q-type sample of by a factor of three. We present the orbital distributions of 64 Q-type near-Earth asteroids, and seek to determine the dominant mechanisms for refreshing their surfaces.  Our sample reveals two important results: i) the relatively steady fraction of Q-types with increasing semi-major axis and ii) the existence of Q-type near-Earth asteroids with Minimum Orbit Intersection Distances (MOID) that do not have orbit solutions that cross Earth. Both of these are evidence that Earth-crossing is not the only scenario by which NEO Q-types are freshened.  The high Earth-MOID asteroids represent 10\% of the Q-type population and all are in Amor orbits.  While surface refreshing could also be caused by Main Belt collisions or mass shedding from YORP spinup, all high Earth-MOID Q-types have the possibility of encounters with Mars indicating Mars could be responsible for a significant fraction of NEOs with fresh surfaces. 

\end{abstract}

%%%%%%%%%%%%%%%%%%%%%%%%%%%%%%%%

\end{frontmatter}

% \linenumbers

%%%%%%%%%%%%%%%%%%%%%%%%%%%%%%%%
\section{Introduction}

Space weathering is a term used to broadly describe the effects of the space environment, such as impacts by high energy particles and micrometeorites, on the surface of airless bodies. The space environment produces a variety of effects on an observed spectrum such as changes in albedo, band depth, and spectral slope, but these effects are not consistent among all bodies. Lunar-style space weathering increases spectral slope and decreases band depth and albedo \citep{Hapke2001,Noble2001,Pieters2000,Taylor2001}. While many laboratory experiments on ordinary chondrites or their mineral constituents reveal space weathering effects similar to lunar-style such as increased spectral slope reddening and decreased albedos \citep{Sasaki2001,Clark2002, Chapman2004, Brunetto2005, Brunetto2006b,Brunetto2006,Brunetto2009}, space weathering on S-complex asteroids as seen by spacecraft missions do not follow lunar-style trends nor do they display weathering trends consistent with each other \citep[e.g.,][]{Veverka1996, Helfenstein1996,Clark2001,Murchie2002,Bell2002,Gaffey2010}. This might not be surprising, however, given that the S-complex encompasses a diverse set of compositions \citep[e.g.,][]{Gaffey1993,Dunn2013}.

A more specific class of asteroids, the Q-type, is currently found primarily (though not exclusively) among Near-Earth Objects (NEOs) that
 are the best spectroscopic matches to LL ordinary chondrites over visible to near-infrared wavelengths  \citep{McFadden1985, 
 Binzel1996}. Because they are a direct meteorite match, they are expected to have undergone processes that disturb their weathered 
 surface regolith, overturning the space weathered grains and revealing fresh, unweathered grains \citep{Binzel2010,Nesvorny2010} on 
 recent timescales \citep[$<$ 1My,][]{Vernazza2009}. 
 The link between asteroid Itokawa and LL ordinary chondrites from the 
 Hayabusa mission as well as the spectral gradient seen from Q-type to S-complex among asteroids, particularly as a function of size, 
 have been considered evidence that increased space weathering changes a spectrum from Q to S 
\citep{Binzel1996,Binzel2001,Binzel2004,Nakamura2011},
 %\citep{Binzel1996,Binzel2004,Binzel2001,Nakamura2011,Thomas2012}, 
 however, considering the compositional diversity of the S-
 complex, the variety of space weathering spectral trends, and recent evidence that some S-complex asteroids are unweathered 
 \citep{MotheDiniz2010}, caution should be taken before generalizing this taxonomic trend to the entire S-complex \citep{Gaffey2010}. 
 In this work we focus on the example of Q-types as markers of unweathered surfaces because there is less compositional variation and 
 thus less ambiguity among that sample, though we note we work under the assumption that all the Q-types are young and fresh.

Multiple mechanisms have been postulated to cause this surface freshening, such as tidal effects from close planetary encounters, YORP spinup, asteroid collisions, and electrostatic levitation of grains from passing through Earth's magnetosphere \citep[e.g.][]{Nesvorny2005,Marchi2006,Binzel2010,Nesvorny2010,Thomas2011, Rivkin2011}. Orbital trends have shown that for the near-Earth asteroid population, planetary encounters play an important role \citep{Marchi2006}. Based on a dataset of 95 objects, \citet{Binzel2010} find that all 20 spectral Q-types in their dataset have an extremely low Minimum Orbit Intersection Distance (MOID) to the Earth. They propose that Earth tidal forces due to close Earth encounters, as suggested by \citet{Nesvorny2005}, are the dominant mechanism for surface refreshing. \citet{Nesvorny2010} note that the orbital distribution of these Q-types is potentially bimodal. Due to the difficulty simultaneously simulating both the low and high semi-major axis groups, they suggest perhaps this bimodality is the result of two different freshening mechanisms.

New observations presented here have more than doubled the number of known Q-types. By examining the orbital distribution of Q-types, we seek to determine the dominant mechanisms for refreshing their surfaces. We calculate Earth and Mars-MOIDs for our sample of 249 S-complex and Q-type NEOs. We investigate a newly revealed group of Q-type objects with Minimum Orbit Intersection Distances (MOID) that do not intersect Earth. We review the potential methods of surface refreshing in the context of the current sample of Q-types, and we explore whether the distribution of these bodies is consistent with theses mechanisms.

\section{Observations}

New observations used in this work are part of the MIT-Hawaii-IRTF joint campaign for NEO  reconnaissance \citep{Binzel2006}, the goal of which is to take routine near-infrared spectral measurements of NEOs. Data from this survey are publicly available at \url{smass.mit.edu}. Observations were taken on the 3-meter NASA Infrared Telescope Facility at the Mauna Kea Observatory. We use the instrument SpeX \citep{Rayner2003}, a near-infrared spectrograph in low resolution mode over 0.8 to 2.5 $\mu$m.

Objects are observed near the meridian (usually $<$ 1.3 airmass) in two different positions (typically denoted A and B) on a 0.8 x 15 arcsecond slit aligned north-south. Exposure times are typically 120 seconds, and we measure 8 to 12 A-B pairs for each object. Solar analogue stars are observed at similar airmass throughout the night.

Reduction and extraction is performed using the Image Reduction and Analysis Facility (IRAF) provided by the National Optical Astronomy Observatories (NOAO) \citep{Tody1993}. Correction in regions with strong telluric absorption is performed in IDL using an atmospheric transmission (ATRAN) model by \citet{Lord1992}. The final spectrum for each object is created by dividing the telluric-corrected asteroid spectrum by the average of the telluric-corrected solar star spectra throughout that night. More detailed information on the observing and reduction procedures can be found in \citet{Rivkin2004} and \citet{DeMeo2008}.

In this work we present 41 Q-type and 23 potential Q-type spectra (this distinction is explained in the next section). Visible data used in this work are from \citet{Bus2002a} and \citet{Binzel2004}. Spectra are presented in the Appendix. The observation dates of these objects are given in Table~\ref{tab:obs}. A full list of all observations is provided in the Supplementary Material.

\begin{table*}
  \caption[]{Observation Table}
 \label{tab:obs}
\centering
\begin{tabular}{rrccccc}
\hline
 \textbf{Number}	&	\textbf{Name}	&	\textbf{Run}	&	\textbf{Observing Date}	& \textbf{Phase Angle}	&	\textbf{V mag}	&	\textbf{Type\footnote{Q is for Q-type, Q: indicates uncertainty between Q- and Sq-type.}}	\\
 \hline
1566	&	Icarus	&	Sp42	&	2005 06 05	&	49	&	16.9	&	Q	\\
1862	&	Apollo	&	Sp48	&	2005 11 22	&	24	&	13.7	&	Q	\\
2212	&	Hephaistos	&	Sp57	&	2006 12 22	&	14	&	15.7	&	Q:	\\
3361	&	Orpheus	&	Sp45	&	2005 10 08	&	6	&	17.3	&	Q	\\
3753	&	Cruithne	&	Sp45	&	2005 10 08	&	57	&	16.6	&	Q	\\
4183	&	Cuno	&	Sp103	&	2011 10 24	&	10	&	16.9	&	Q	\\
4688	&	1980WF	&	Sp03	&	2001 01 29	&	43	&	17.9	&	Q	\\
5143	&	Heracles	&	Sp55	&	2006 10 25	&	16	&	15.3	&	Q	\\
5660	&	1974MA	&	Sp43	&	2005 07 09	&	37	&	15.9	&	Q	\\
7336	&	Saunders	&	Sp93	&	2010 09 06	&	12	&	16.6	&	Q	\\
7341	&	1991VK	&	Sp103	&	2011 10 24	&	25	&	16.3	&	Q	\\
11054	&	1991FA	&	Sp09	&	2002 01 12	&	43	&	15.8	&	Q:	\\
19764	&	2000NF5	&	Sp93	&	2010 09 06	&	19	&	16.2	&	Q:	\\
23183	&	2000OY21	&	Sp49	&	2006 01 30	&	33	&	15.5	&	Q:	\\
23187	&	2000PN9	&	Sp05	&	2001 03 28	&	51	&	16.6	&	Q:	\\
39572	&	1993DQ1	&	Sp73	&	2008 09 02	&	65	&	15.6	&	Q:	\\
66146	&	1998TU3	&	Sp74	&	2008 10 02	&	70	&	15.5	&	Q	\\
85236	&	1993KH	&	Sp76	&	2008 12 03	&	61	&	16.4	&	Q:	\\
85839	&	1998YO4	&	Sp89	&	2010 03 16	&	22	&	16.3	&	Q:	\\
88254	&	2001FM129	&	Sp89	&	2010 03 17	&	81	&	15.7	&	Q	\\
89958	&	2002LY45	&	Sp98	&	2011 04 05	&	10	&	16.3	&	Q	\\
136923	&	1998JH2	&	Sp102	&	2011 09 25	&	45	&	17.0	&	Q:	\\
137032	&	1998UO1	&	Sp74	&	2008 10 02	&	54	&	13.8	&	Q	\\
138883	&	2000YL29	&	Sp84	&	2009 09 20	&	36	&	15.4	&	Q	\\
139622	&	2001QQ142	&	Sp58	&	2007 01 21	&	55	&	16.9	&	Q	\\
143487	&	2003CR20	&	Sp84	&	2009 09 20	&	51	&	16.5	&	Q	\\
143651	&	2003QO104	&	Sp79	&	2009 03 30	&	41	&	16.0	&	Q	\\
152931	&	2000EA107	&	Sp89	&	2010 03 16	&	50	&	16.7	&	Q	\\
154244	&	2002KL6	&	Sp81	&	2009 06 21	&	48	&	15.5	&	Q:	\\
154715	&	2004LB6	&	sp96	&	2011 01 06	&	51	&	17.7	&	Q:	\\
162058	&	1997AE12	&	Sp26	&	2003 10 16	&	59	&	17.0	&	Q	\\
162117	&	1998SD15	&	Sp64	&	2007 10 02	&	32	&	16.8	&	Q:	\\
162483	&	2000PJ5	&	Sp82	&	2009 08 08	&	53	&	17.3	&	Q	\\
163081	&	2002AG29	&	Sp102	&	2011 09 25	&	49	&	17.2	&	Q	\\
163697	&	2003EF54	&	Sp83	&	2009 08 24	&	48	&	16.1	&	Q	\\
164400	&	2005GN59	&	Sp73	&	2008 09 02	&	27	&	15.7	&	Q:	\\
184266	&	2004VW14	&	Sp69	&	2008 04 13	&	53	&	16.3	&	Q:	\\
206910	&	2004NL8	&	Sp78	&	2009 03 02	&	16	&	16.9	&	Q	\\
218863	&	2006WO127	&	Sp89	&	2010 03 17	&	66	&	16.0	&	Q	\\
219071	&	1997US9	&	Sp104	&	2011 10 31	&	23	&	16.9	&	Q	\\
274138	&	2008FU6	&	Sp99	&	2011 04 30	&	52	&	16.5	&	Q	\\
	&	1998SJ70	&	Sp74	&	2008 10 02	&	7	&	16.6	&	Q	\\
	&	2002NY40	&	Sp16	&	2002 08 17	&	20	&	11.1	&	Q	\\
	&	2003FH	&	Sp104	&	2011 10 31	&	65	&	16.6	&	Q	\\
	&	2003MJ4	&	Sp92	&	2010 07 11	&	35	&	17.3	&	Q	\\
	&	2003UV11	&	Sp94	&	2010 10 13	&	23	&	18.1	&	Q	\\
	&	2004QJ7	&	Sp103	&	2011 10 24	&	45	&	17.5	&	Q	\\
	&	2005ED318	&	Sp40	&	2005 05 11	&	59	&	16.4	&	Q	\\
	&	2006VB14	&	Sp76	&	2008 12 03	&	52	&	16.4	&	Q	\\
	&	2007LL	&	Sp71	&	2008 06 11	&	19	&	17.3	&	Q	\\
	&	2007RU17	&	Sp94	&	2010 10 14	&	4	&	16.1	&	Q	\\
	&	2008CL1	&	Sp68	&	2008 03 10	&	26	&	16.1	&	Q	\\
	&	2008HS3	&	Sp70	&	2008 05 10	&	50	&	17.5	&	Q	\\
	&	2011OV4	&	Sp101	&	2011 08 22	&	15	&	17.2	&	Q	\\
	&	2011PS	&	dm03	&	2011 09 03	&	39	&	16.9	&	Q	\\
	&	2000QW7	&	Sp01	&	2000 09 04	&	29	&	13.7	&	Q:	\\
	&	2002GO5	&	Sp12	&	2002 04 14	&	11	&	14.6	&	Q:	\\
	&	2006HQ30	&	Sp51	&	2006 06 11	&	32	&	15.7	&	Q:	\\
	&	2006VQ13	&	Sp56	&	2006 11 21	&	31	&	16.5	&	Q:	\\
	&	2007DT103	&	Sp62	&	2007 07 31	&	86	&	14.3	&	Q:	\\
	&	2008FU6	&	Sp98	&	2011 04 05	&	4	&	16.6	&	Q:	\\
	&	2008UE7	&	Sp76	&	2008 12 03	&	24	&	16.1	&	Q:	\\
	&	2010CM44	&	Sp90	&	2010 04 16	&	8	&	16.7	&	Q:	\\
	&	2010LY63	&	Sp93	&	2010 09 07	&	38	&	15.5	&	Q:	\\
  \hline

\end{tabular}
\end{table*}

\section{Classification of near-IR data}
\subsection{Method of Classification}

\begin{figure}
  \centering
  \includegraphics[width=0.5\textwidth]{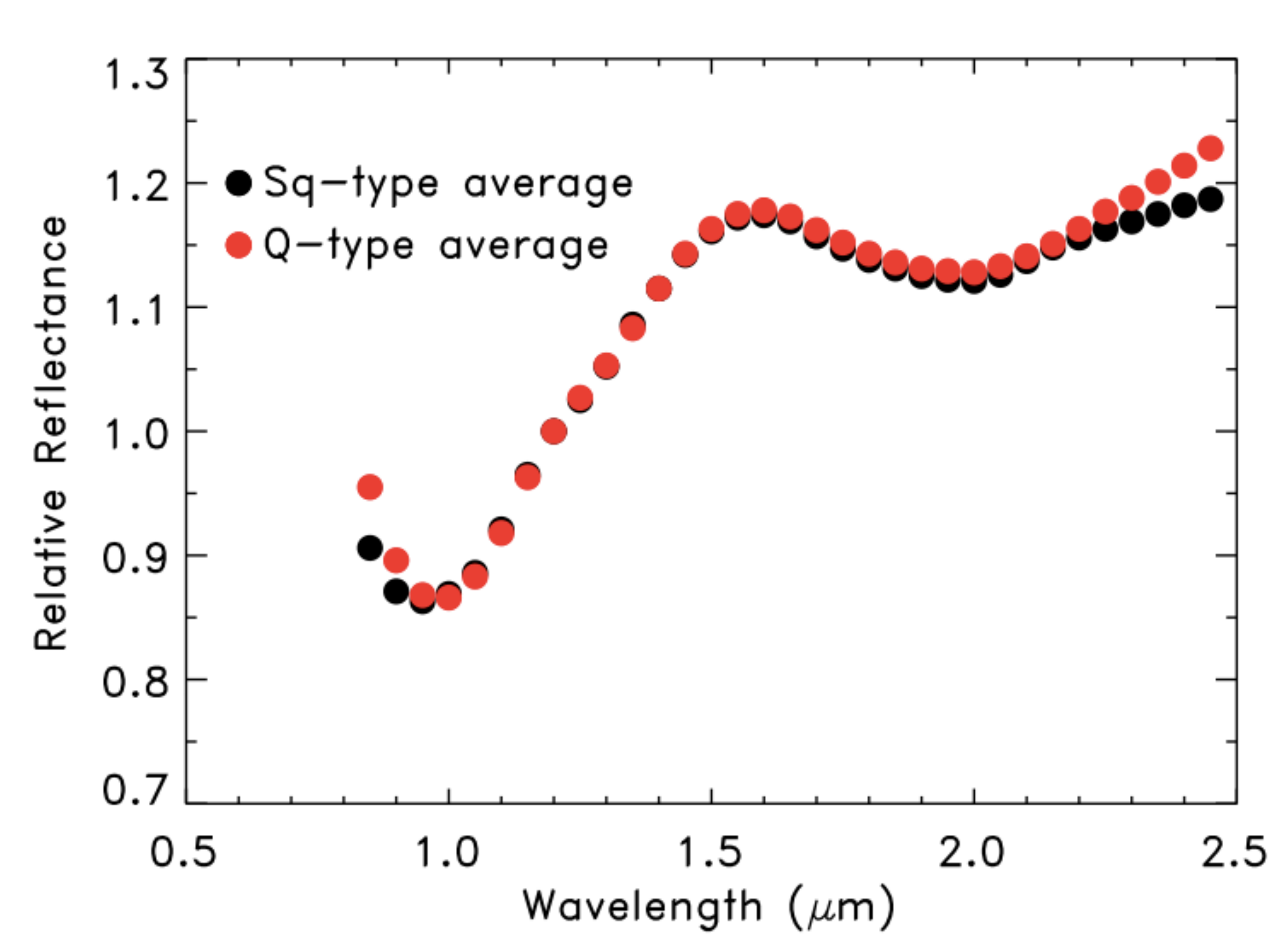}
  \caption[]{%
    Comparison of Bus-DeMeo average Sq-type (red) and average Q-type (blue) as revealed over the 0.85 to 2.45 $\mu$m range and normalizing the data at 1.25 $\mu$m. The only significant difference between these spectra is the slope between 0.85 and 0.95 $\mu$m. Over all other wavelengths the values are the same within 1$\sigma$ errors.} 
  \label{fig: qsqcomp}
\end{figure}

Objects that have \textit{both} visible \textit{and} near-infrared spectral measurements can in most cases be easily and unambiguously classified using the Bus-DeMeo taxonomy and the online classifier  \citep[smass.mit.edu/busdemeoclass.html,][]{DeMeo2009}. Distinguishing Q-types is more challenging in the near-infrared because weathering effects are more prominent in the visible wavelength range (See Fig~\ref{fig: qsqcomp}). When only the near-infrared portion of the spectrum is available, the Bus-DeMeo taxonomy does not have a unique way to distinguish among S-complex and Q-type using Principal Component Analysis.

To distinguish Q-types from S-, Sr-, and Sq-types in the near-infrared range we use 6 parameters: the 1 $\mu$m band minimum, a modified version of the 1 $\mu$m band width, depth, and area, the ``potential'' 1 $\mu$m band depth, and the slope of the 1 $\mu$m ``tail''  (TSlope).  The band minimum (B$_{Min}$) is defined as the minimum value of a gaussian fit to the 1 $\mu$m band. TSlope is the slope of the data from 0.85 to 0.95 and represents the steepness of the short-wavelength wall of the 1 $\mu$m band, which gives an indication of the characteristics in the visible range. 

\begin{figure}
  \centering
  \includegraphics[width=0.5\textwidth]{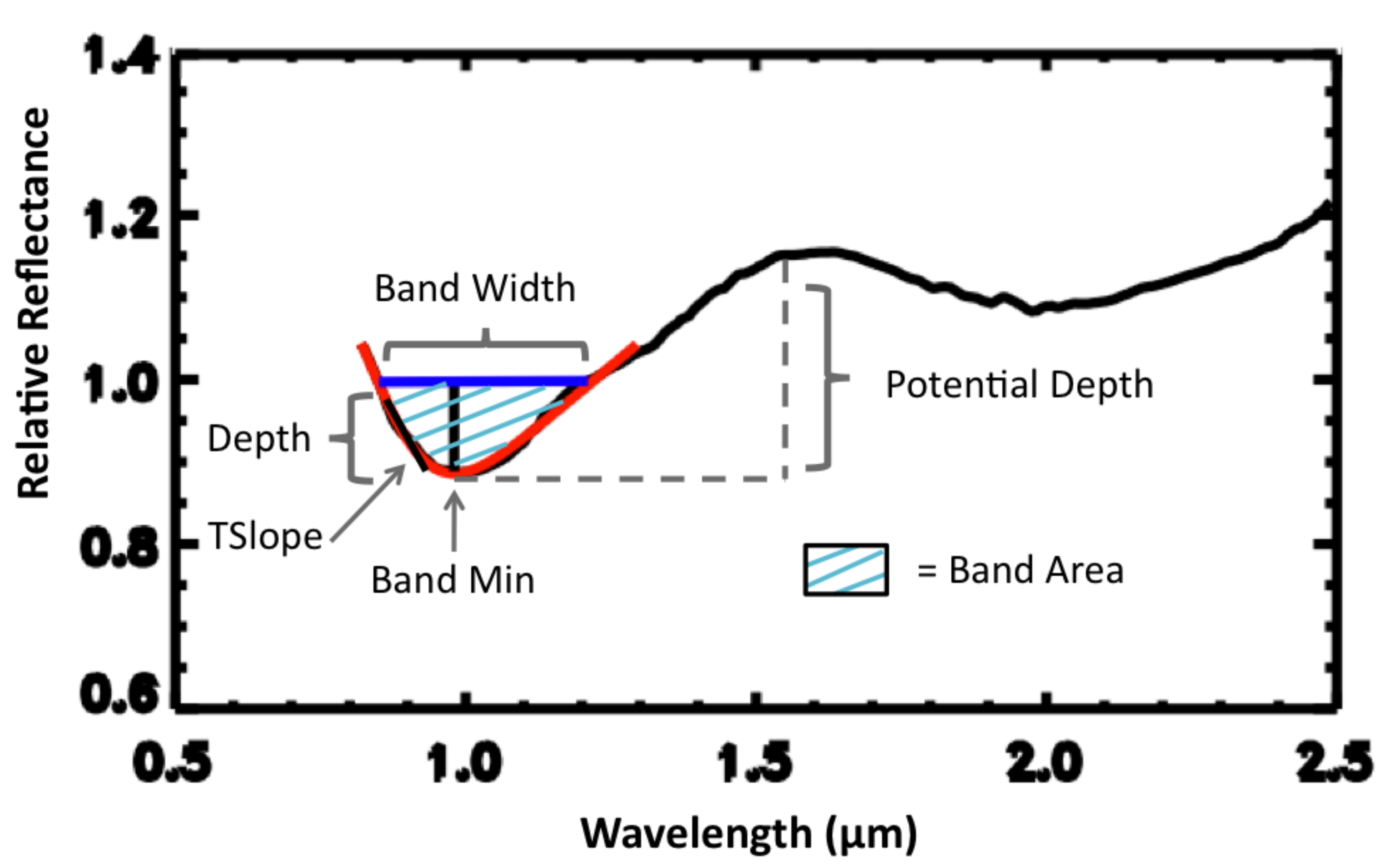}
  \caption[]{%
    	Description of each parameter, defined in Section 3, used to distinguish Q-types from S-complex spectra when only near-infrared data (0.85 to 2.45 $\mu$m) are available.} 
  \label{fig: bandparam}
\end{figure}

Because the spectral range does not cover the entire 1 $\mu$m band, we define a modified version of band parameters meant to be internally consistent, but not useful for mineralogy or direct comparison with other work. The spectral band parameters we define here are different from those used in prior studies (e.g., Cloutis et al. 1986, Gaffey, 1997, Thomas et al. 2010). Herein, the ``Band Depth'' (B$_{Depth}$) is defined as the vertical distance between reflectance data at 0.85 $\mu$m and the band minimum. The ``Potential Band Depth'' (P$_{Depth}$) is the expected maximum depth we would expect the band to have (the case where the continuum slope is zero over the width of the band) and is calculated as the vertical distance between the band minimum and the continuum level at the end of the band defined here as the maximum value between 1.45 and 1.55 $\mu$m. The ``Band Width'' (B$_{Width}$) is defined as the horizontal distance between the reflectance at 0.85 and the same reflectance value on the long-wavelength end of the band (B$_{End}$). To calculate the ``Band Area'' (B$_{Area}$) we compute the definite integral of the reflectance spectrum over the horizontal distance defined by the Band Width (which yields the area underneath the band) and subtract that from the area underneath the horizontal line defining the Band Width. Figure \ref{fig: bandparam} visually depicts each parameter on a sample spectrum. The equations defining these parameters are as follows:

\begin{equation}
B_{Depth} = y(0.85 \mu m) - y(B_{Min})
\end{equation}
\begin{equation}
P_{Depth} = max(y(1.45:1.55 \mu m)- y(B_{Min}))
\end{equation}
\begin{equation}
X_{B_{End}}=x value where(y(X_{B_{End}})=y(0.85  \mu m))
\end{equation}
\begin{equation}
B_{Width} = X_{B_{End}} -0.85  \mu m
\end{equation}
\begin{equation}
B_{Area} = B_{Width} * y(0.85 \mu m) - \int\limits_{x=0.85}^{X_{B_{End}}} y dx 
\end{equation}

\begin{figure}
  \centering
  \includegraphics[width=0.5\textwidth]{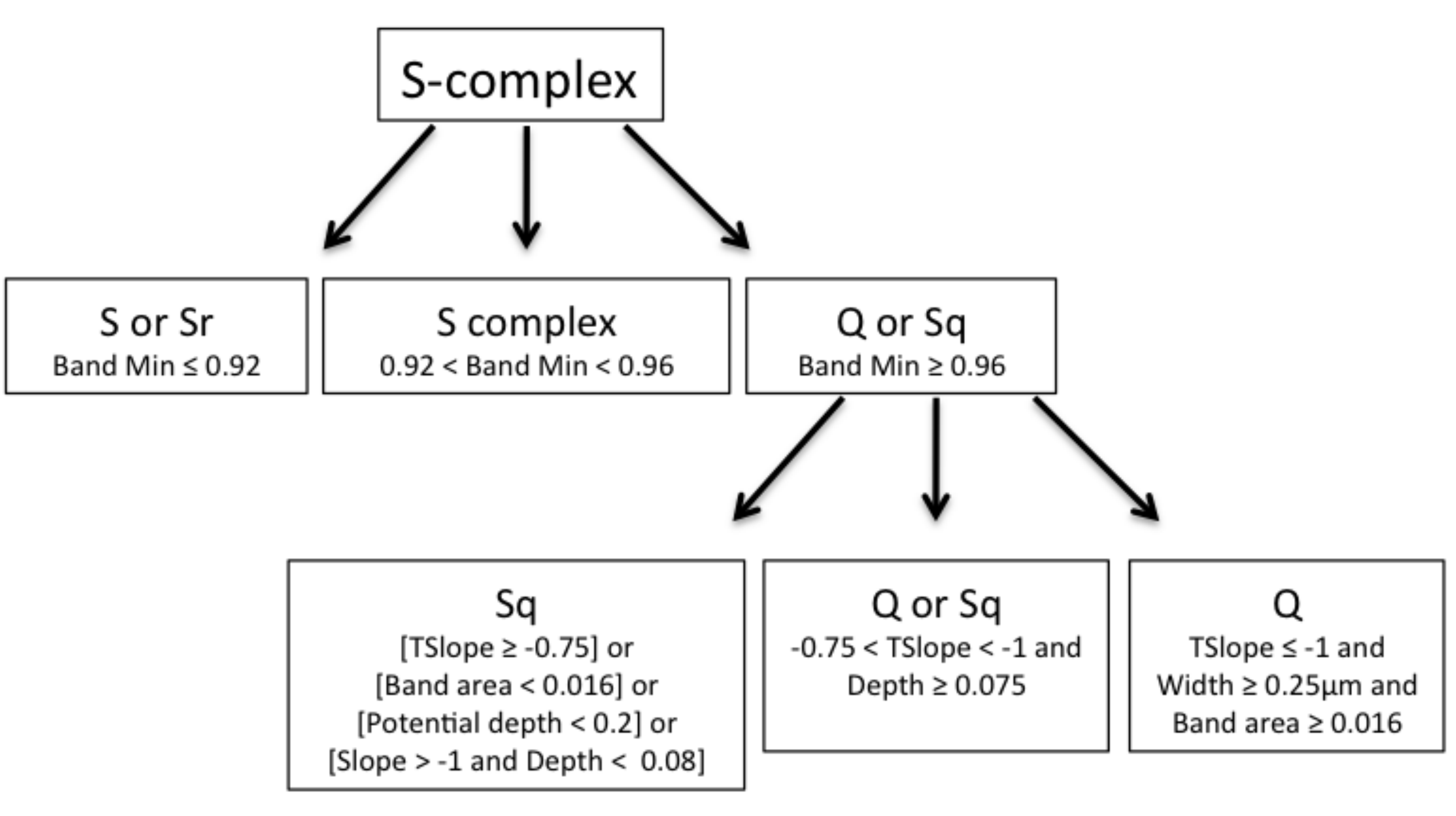}
  \caption[]{%
    	Flow chart of boundaries for band parameters defined in Section 3 to categorize near-infrared S-complex spectra used in this work.} 
  \label{fig: flowchart}
\end{figure}

To best distinguish the spectral type we set boundaries using the data that included visible and near-infrared ranges (and thus could formally be classified). We find that generally as the B$_{Min}$, TSlope, B$_{Depth}$, B$_{Width}$, and B$_{Area}$ increase, the classification trends from S- and Sr-type to Sq-type and then to Q-types. The P$_{Depth}$ parameter is important to identify the few objects with wide but very shallow bands from being confused with a potential Q-type classification.  We separate the objects into the following categories: S, S:, Sq, Q:, and Q, where S includes the entire S complex, S: includes S-type and Sr-type spectra, Q: includes Q-type and Sq-type spectra, and Sq and Q are analogous to the classes defined in \citet{DeMeo2009}. A flowchart of the classification method including the values for each parameter for each class is given in Fig.~\ref{fig: flowchart}
 
We compare classifications using this flowchart with objects that are ``formally'' classified with both visible and near-infrared data, to verify that the classification is reasonable.  For all formal Q-types 70\% (9/13) fell in either the Q-type or Q:-type (uncertain between Sq- or Q-type) category. 15\% of formal Q-types fell in each of the Sq- and S-complex categories. Among formal Sq-types, 5, 40, 30, and 25\% fell in the Q:, Sq, S complex, and S: categories, respectively.  Looking at the values conversely, of all the objects classified as Q, 100\% (3 objects) were formally defined Q-types. Of the objects whose class was deemed Q:, 6 were formal Q-types and 2 were formal Sq-types, suggesting any object falling within this ``Q or Sq'' bin has a 75\% chance of being a Q-type. However, to keep our identifications conservative, in this work we assign any object falling in the Q: category a 50\% probability of being Q-type. For objects classified as Sq there were 3, 20, 4, and 1 objects for formal Q-, Sq-, S-, and Sr-types, respectively, showing that 70\% of objects in this bin are actually Sq-types. A table of results for band parameters and classification for the entire dataset is available in the Supplementary Material.

\subsection{Effects of phase angle on classification}
There are host of different properties and effects that can change the spectral parameters we measure and can mimic the effects of space weathering. These include phase angle, composition, grain size, temperature and observing conditions \citep[e.g.,][]{Reddy2012}. While it is not possible to correct for all these causes we must still consider the impact they would have on the spectra and the overall results. To narrow the dataset compositionally, we also inspect trends not only for the full dataset, but also for a subsample that does not include Sr-types, which are compositionally inconsistent with Q-types because of their more narrow 1 micron absorption features. There are many other compositional parameters to be considered that are beyond the scope of this work.

The phase angle at which an asteroid is observed can have an important effect on spectral features such as the depth of the one micron band and the slope of the spectrum \citep[e.g.,][]{Sanchez2012,Clark2002b}. NEOs are observed at a wide variety of phase angles, and although we do not correct for phase angle in this work, we caution it could have an effect on the asteroid classification. We find that some of the trends in our overall sample with phase angle (band minimum and band width) would cause an object to be more likely classified as an S-type, while other parameters (Tslope and band depth) are more likely to push the classification toward Q-type. In our sample, we do find relatively more Q-types at higher phase angles than other classes, however, if phase angle was causing the classification toward more Q-like spectra then we would also expect there to be an overabundance of Sq-types and Q:-types at higher phase angles relative to S-types which is not seen.

Among asteroids in our sample that are observed multiple times at different phase angles we find no correlation between class and phase angle. Larger phase angles do not correlate with a change from S to Sq to Q.
Roughly half of these objects have a Q-type spectrum as the lower phase angle observation and half are the higher phase observation. It is possible that the effects of phase angle are diminished by the many other factors that can affect a spectrum such as data quality and other observing conditions. Throughout this work, however, when we make a claim based on our sample, such as detecting Q-types that do not cross Earth, we also inspect a subsample of our data taken only at low phase angles to be certain we are not misinterpreting the results.

\section{Results}
\begin{figure}
  \centering
  \includegraphics[width=0.5\textwidth]{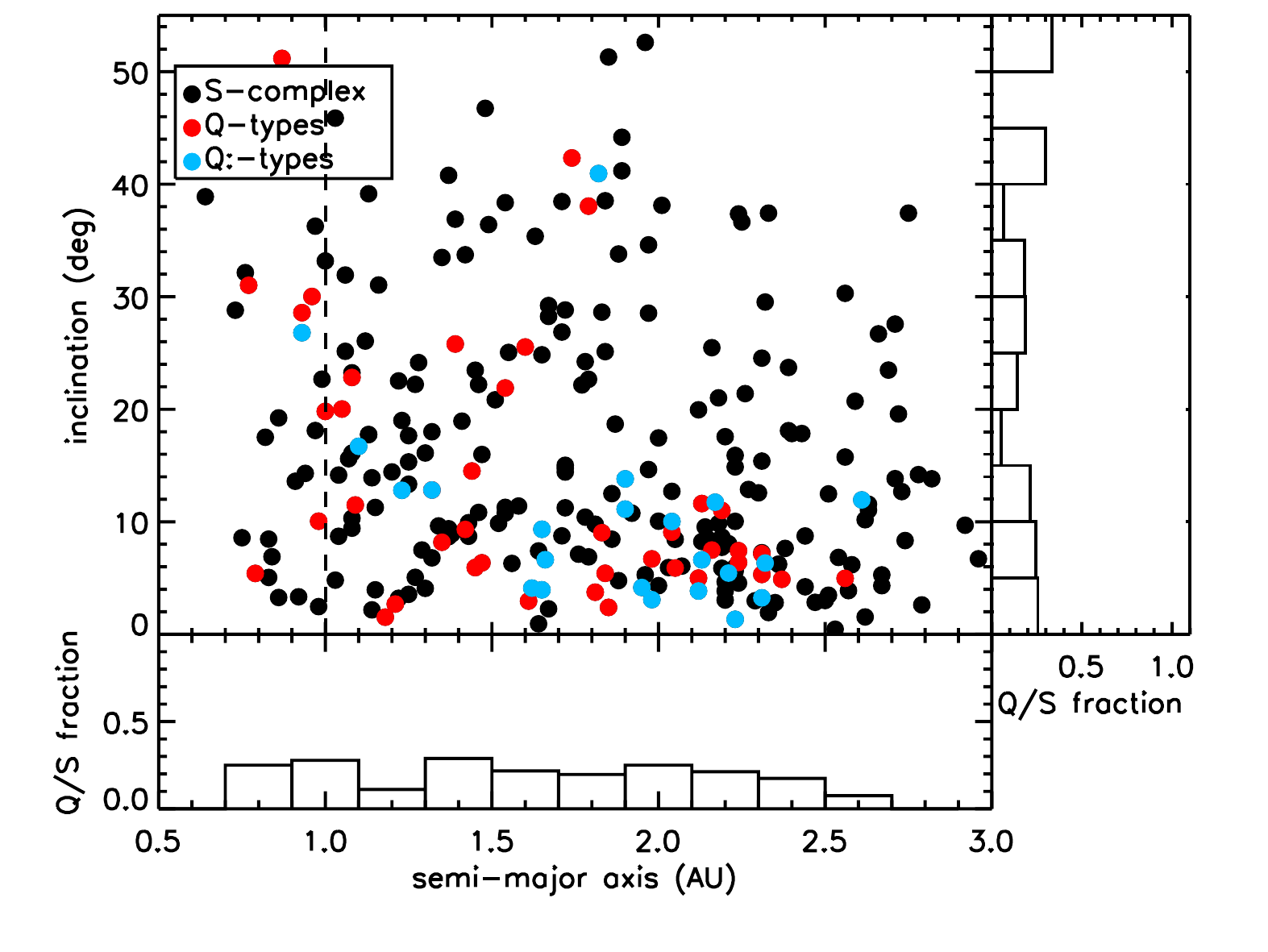}
  \caption[]{%
    	The orbital distribution of S-complex and Q-type near-Earth asteroids in semi-major axis and inclination. 
	Red dots are Q-type asteroids, blue dots are Q: and black dots are S-complex. The fraction of Q to 
	S is shown for semi-major axis in the lower histogram and for inclination in the side histogram.
	A weighting of 0.5 is given to ``Q:'' (degenerate between Sq and Q) objects in the histograms.
	The distribution in semi-major axis is relatively constant until about 2.4 AU aside from the smaller
	Q-type fraction in the 1.1 to 1.3 AU bin, although this difference is not statistically significant.
	}
	% It is evident in this plot that at low inclinations (i $<$ 15 degrees) Q-type asteroids are more 
	%frequent at higher semi-major axes than they are for orbits closer to the Earth.} 
  \label{fig: ai}
\end{figure}

We identify 41 Q-types and 23 Q:-types (defined here as falling in the Sq/Q bin and assigned a 50\% probability) in the NEO population.  This means that statistically we have 52.5 Q-types in our sample (41 + $\frac{23}{2}$). In Fig.~\ref{fig: ai} we plot semi-major axis versus inclination for S-complex and Q-type near-Earth asteroids. While in this work we use the S-complex as a baseline for comparison with Q-type orbital distributions, we must stress that the S-complex is much more compositionally diverse than the Q-type subset and displays a variety of weathering trends that complicate our understanding of the full population. We find that the fraction of Q-types to S-complex objects is generally constant for all semi-major axes out to about 2.4 AU. There is a deficit of Q-types in the 1.1 to 1.3 AU bin compared to surrounding bins. This difference is not statistically significant, although it is intriguing that the deficit lies in a region very close to the Earth where we would expect planetary encounters to be more frequent than for larger semi-major axes. 

If we restrict our sample to only low phase angle observations ($<$30 degrees), we find the fraction of Q-types as a function of distance is still relatively flat as seen before. The difference between the full and low phase samples is that we now see slightly lower relative fractions at lower semi-major axis. This could be that we typically observe these objects at higher phase angles and were over-estimating Q-types in this sample, however, there is no significant difference between the typical phase angle for observations of Aten, Apollos, and Amors in our sample, so this may not be the case. Our sample size is smaller with this restricted sample, but if this difference is true, then it actually strengthens the case that other mechanisms play an important role in refreshing because objects with higher semi-major axis have longer orbital periods and thus approach Earth's orbit less frequently than those with lower semi-major axis.

%%%%-- BEGIN --%%%%%%%%%%--- Plot ---%%%%%%%%%%%%%%%%%%%%%%%%%%%%%%
%%%%--  END  --%%%%%%%%%%--- Plot ---%%%%%%%%%%%%%%%%%%%%%%%%%%%%%%

\begin{figure}
  \centering
  \includegraphics[width=0.5\textwidth]{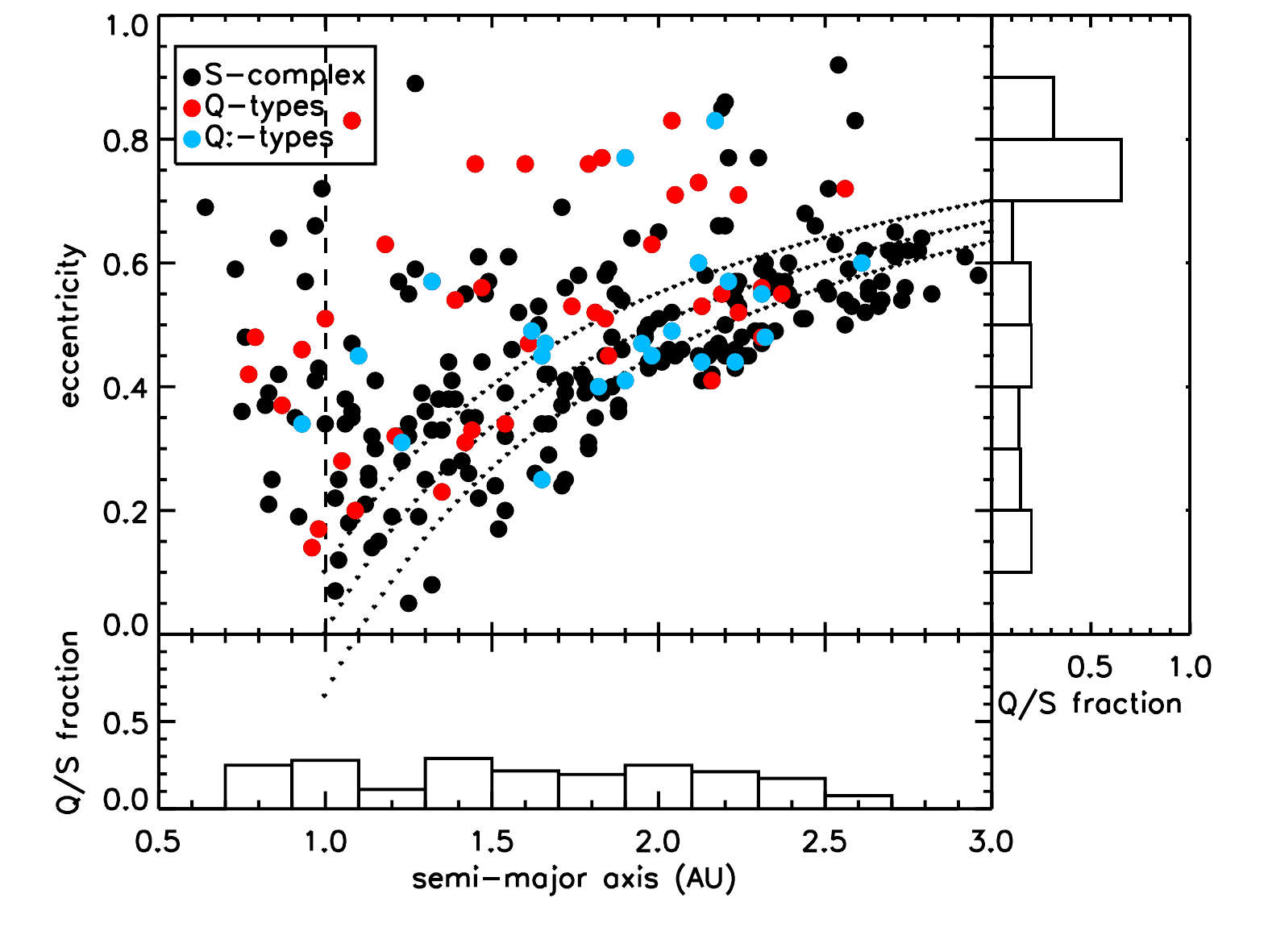}
  \caption[]{%
    	The orbital distribution of S-complex and Q-type near-Earth asteroids for semi-major axis and eccentricity. 
	Red dots are Q-type asteroids, blue dots are Q: and black dots are S-complex. The fraction of Q to 
	S is shown for semi-major axis in the lower histogram and for eccentricity in the side histogram.
	A weighting of 0.5 is given to ``Q:'' (degenerate between Sq and Q) objects in the histograms.
	Of note is the high fraction of Q-types at high eccentricity, although the sample size is too
	small to draw any conclusions.	}
  \label{fig: ae}
\end{figure}

Figure~\ref{fig: ae} shows the same data plotted according to eccentricity. There is a larger fraction of Q-types at higher eccentricities. While this could have important implications for freshening mechanisms, we note that the sample size in each bin is much smaller for higher eccentricities than lower, weakening the robustness of this trend.

\begin{figure}
  \centering
  \includegraphics[width=0.5\textwidth]{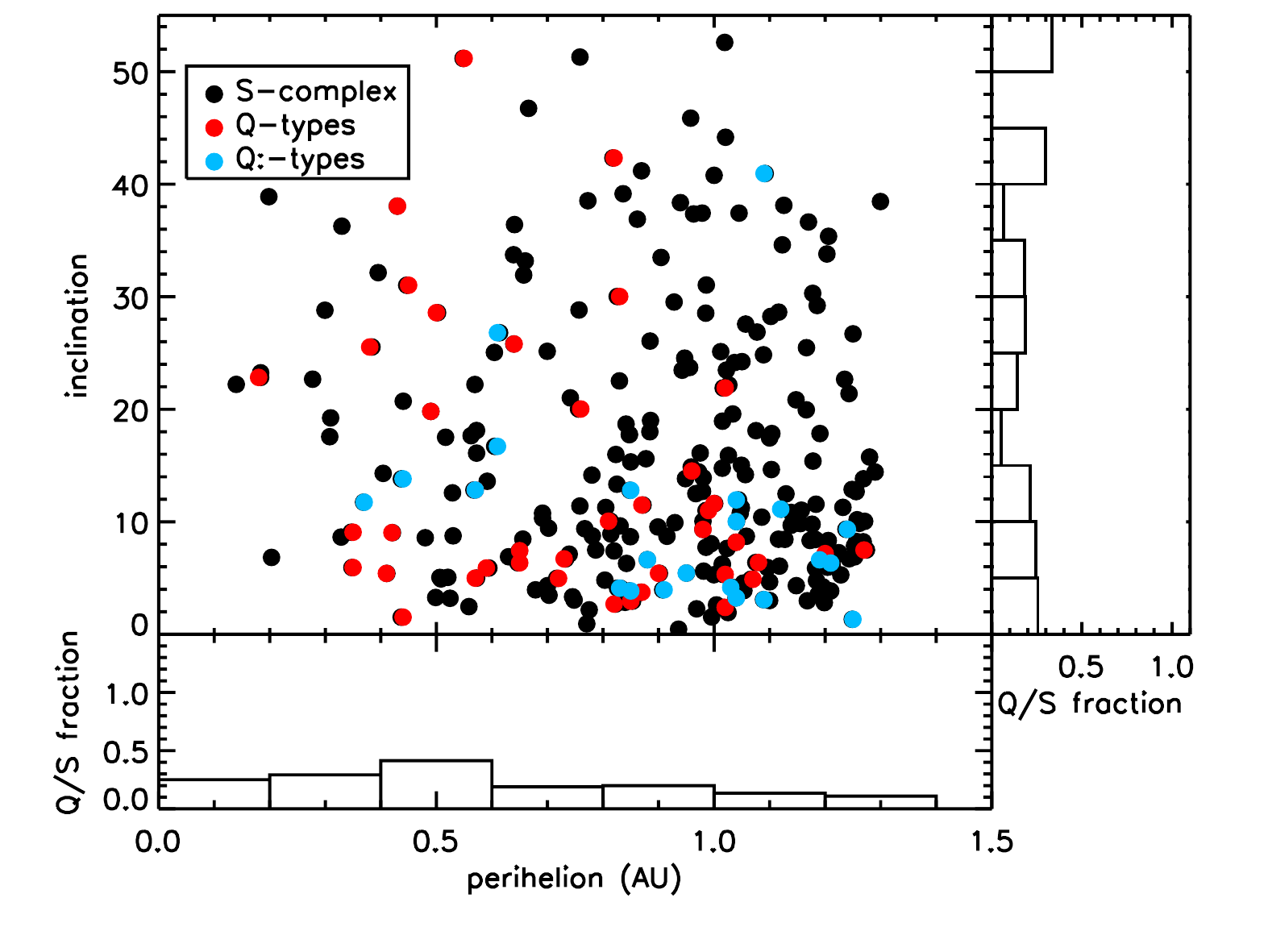}
  \caption[]{%
    	The orbital distribution of S-complex and Q-type near-Earth asteroids for perihelion and inclination. 
	Red dots are Q-type asteroids, blue dots are Q: and black dots are S-complex. The fraction of Q to 
	S is shown for perihelion in the lower histogram and for inclination in the side histogram.
	A weighting of 0.5 is given to ``Q:'' (degenerate between Sq and Q) objects in the histograms.
	The relative fraction of Q-types peaks at 0.5 AU. There is a small sample of Q-types with high 
	perihelia ($>$1.2AU) suggesting they are decoupled from the Earth.	} 
  \label{fig: qi}
\end{figure}

In Fig.~\ref{fig: qi} we plot perihelion versus inclination for S-complex and Q-type near-Earth asteroids.  The fraction of Q-types is highest at a perihelion near 0.5 AU. Q-types also appear more abundant with perihelia smaller than 0.5 compared to larger, however, we again note that the number of objects is low in the bins with more extreme orbits.  There are six Q: objects that have orbits with perihelia near 1.2 AU.  Their current orbits do not put them on an Earth-crossing path, however, NEO orbits experience ample perturbations which radically alter their orbits on short time scales. To explore the potential orbital histories of our entire sample (249 objects), we calculate the orbit and Minimum Orbit Intersection Distance (MOID) to Earth and to Mars over the past 500,000 years. This is done in the same manner as \citet{Binzel2010} using the \texttt{swift\_rmvs3} code from \citet{Levison1994} with a 3.65-day timestep and output values computed at 50-year intervals, accounting for the eight planets Mercury to Neptune. Our integrations for each asteroid included six additional clones, test particles with the same initial position, offset with velocities differing by $\pm 6 x 10^{-6}$ astronomical units (AU) per year in each Cartesian component. These clones are important to evaluate the range of potential orbital evolutions. We define a ``close approach'' as a MOID smaller than the Earth-moon distance.

We do not calculate Venus-MOIDs in this work. Venus is likely roughly as effective at refreshing the surface of an asteroid as Earth because of its similar size and mass though fewer NEAs have Venus-crossing orbits. However, any NEA that is Venus-crossing is also Earth-crossing so for this work we do not consider the effect of Venus since we cannot distinguish encounters with Venus from those with Earth.

\begin{figure}
  \centering
  \includegraphics[width=0.5\textwidth]{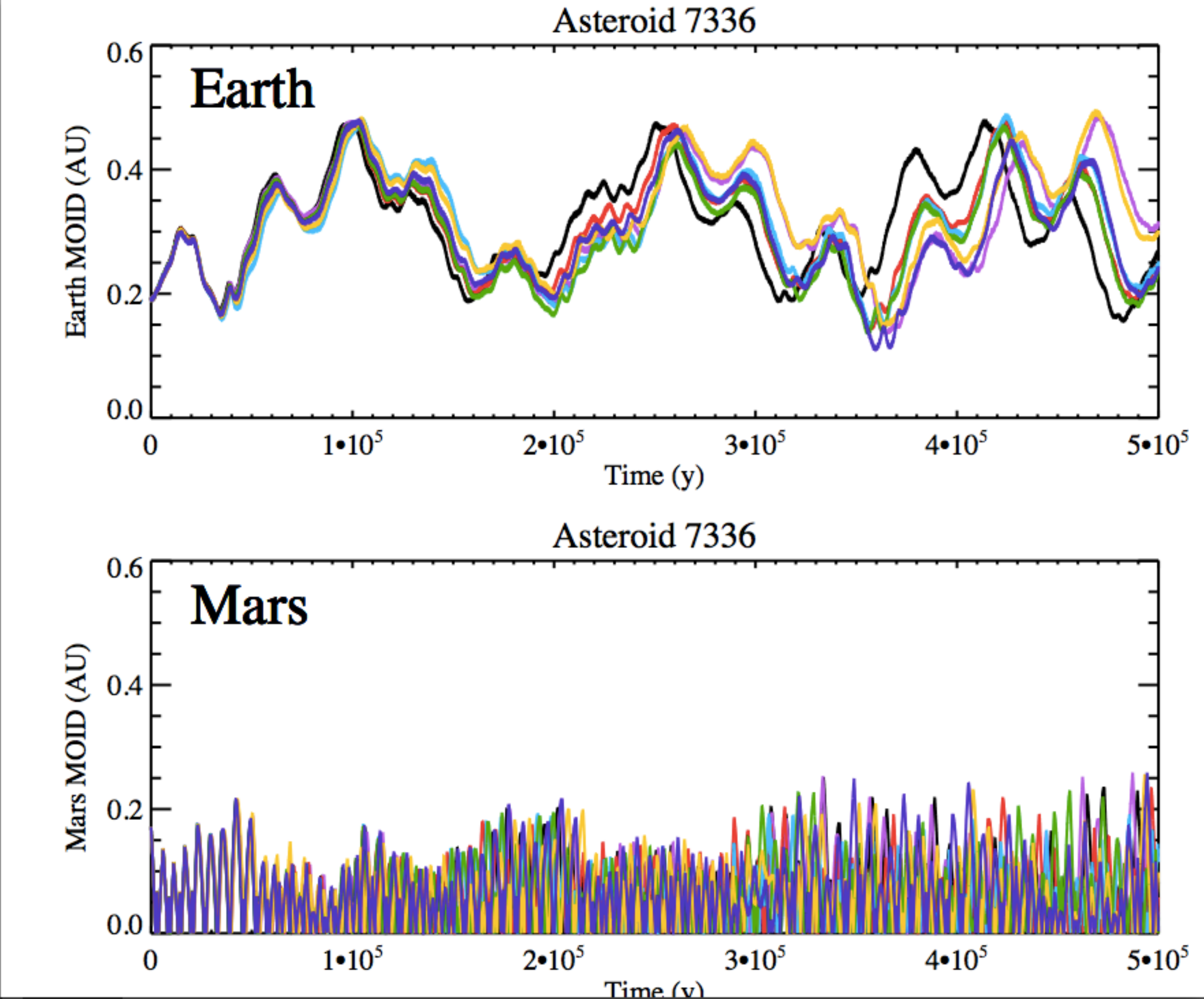}
  \caption[]{%
	Minimum Orbit Intersection Distance (MOID) calculations for Q-type asteroid (7336) and Earth (top) and Mars (bottom).
	This asteroid and its 6 clones did not pass close to the Earth in the past 500,000 years
	according to the simulation, suggesting that there is a low probability that Earth encounter was responsible
	for freshening its surface. However, this asteroid and its 6 clones frequently
	pass very close to Mars suggesting that Mars encounters have the capability to  
	freshen its surface.  } 
  \label{fig: moid-earth}
\end{figure}

The MOID calculations reveal 7 interesting high Earth-MOID objects, i.e. objects showing no close Earth encounters in our 500,000 year integration. These are (7336) Saunders, (138883) 2000YL$_{29}$, and 2011 OV$_4$ that are confident Q-types which we note were observed at phase angles of 12, 36, and 15 degrees, respectively, and (19764) 2000 NF$_5$, (23183) 2000 OY$_{21}$, (85839) 1998 YO$_4$, (136923) 1998 JH$_2$ that are possibly Q-types (Q:)  and were observed at phase angles of 19, 33, 22, and 45 respectively.  Five of the seven have low inclinations (i$<$10 degrees) and high perihelia (q$\gtrsim$1.2), and four of the seven have semi-major axes within the inner main belt (2.0 $<$ a $<$ 2.5 AU).  All are in Mars-crossing Amor orbits.  (7336) has both visible and near-infrared spectral data revealing unambiguously its Q-type classification and presumably its fresh surface. We plot Earth- and Mars-MOIDs for this object in Figure~\ref{fig: moid-earth}.  All 7 of these objects that do not cross Earth have orbits that are expected to regularly and frequently experience deep encounters with Mars according the the Mars-MOID calculations.  

\citet{Binzel2010} found that all 20 of the Q-types in their sample had low Earth-MOIDs. Of their entire sample, an asteroid had a 78\% chance of being on a close Earth-crossing orbit. Using binomial statistics they find with 99.1\% confidence that the Q-type MOID distribution is not random. In our sample, 81.2\% of the asteroids have low Earth-MOID values. Of our Q-type sample 5 out of 52.5 (the Q: objects are half weighted) have high MOID values (9.5\%). There is a 96.8\% chance that this is non-random using binomial statistics. However, if we remove Sr-types from the sample (Sr-types are not considered mineralogically equivalent to Q-types due to their narrow 1 $\mu$m band) there is then a 91.6\% chance that the outcome is non-random. The probability that all Q-types are caused \textit{only} by close encounters with Earth is less robust with this larger sample, illustrating the likelihood of an additional mechanism at work.

%%%%-- BEGIN --%%%%%%%%%%--- Plot ---%%%%%%%%%%%%%%%%%%%%%%%%%%%%%%
%%%%--  END  --%%%%%%%%%%--- Plot ---%%%%%%%%%%%%%%%%%%%%%%%%%%%%%%

%%%%-- BEGIN --%%%%%%%%%%--- Plot ---%%%%%%%%%%%%%%%%%%%%%%%%%%%%%%
%%%%-- BEGIN --%%%%%%%%%%--- Plot ---%%%%%%%%%%%%%%%%%%%%%%%%%%%%%%
%%%%--  END  --%%%%%%%%%%--- Plot ---%%%%%%%%%%%%%%%%%%%%%%%%%%%%%%

\section{Discussion}
%Read Bottke birth of NEO + Earth/Mars crossing population size ratios
%Nesvo YORP/arguments against other methods
%Pravec and yorp effiiciency --> yorp efficiency per distance/ high inclinations
%Binz - how does new data change rate/distance?

The relatively constant fraction of Q-types to S-complex objects according to semi-major axis among NEOs seen here (see Figs.~\ref{fig: ai} and \ref{fig: ae}) suggests that the bimodality seen in previous work \citep{Binzel2010,Nesvorny2010} was likely due to the small sample size. However, this semi-major axis trend is surprising if Earth and Venus encounters are the sole mechanism for surface freshening. If this were the case we would expect the fraction of Q-types to be highest between 0.7 and 1.0 AU and to drastically taper off with increasing semi-major axis as the probability of encounter decreases \citep[see ][ for orbital simulations]{Nesvorny2010}.  Evidence for alternative freshening mechanisms is strengthened by the presence of a newly discovered population of objects with large Earth-MOIDs.  Particularly interesting are the four objects with large perihelia ($\gtrsim$ 1.2 AU) and large semi-major axes (a $>$ 2.0). 

Here we further address dynamically why these high Earth-MOID, high-perihelia, high semi-major axis objects are likely to have never encountered Earth or Venus. While close encounters with Earth or Venus can cause significant changes in the orbit, such as large changes in semi-major axis or eccentricity, perhelion is largely unchanged because the Tisserand parameter is conserved and the body follows a random walk along the contour of Tisserand invariance \citep{Tisserand1882,Greenberg1993}.  Because these objects are Amors with semi-major axes greater than 2 AU, their orbits are consistent with recent escape from the main belt via the $\nu_6$ resonance \citep{Bottke2005,Morbidelli2002}. Objects recently ejected from the main belt often enter the NEO population first as an Amor. However, it is very difficult for an older, more evolved NEO to change from an Aten or Apollo orbit back out to an Amor because of the higher eccentricity of an evolved NEO \citep{Morbidelli2002}.  NEOs with a $>$ 2 AU are considered relatively \textit{dynamically} ``young'' because they haven't evolved into lower semi-major axis orbits through planetary encounters \citep{Morbidelli2002}.  The average lifetime of a $>$ 2 AU is $\sim$ 1 M.y. while for evolved orbits with a $<$ 2 AU the average lifetime jumps to $\sim$ 10 M.y.  \citep{Gladman1997,Gladman2000}. 

Thus the Amor orbit, and high perihelia and semi-major axes suggest that these objects have never yet been on Earth or Venus encountering orbits. The Earth-MOID calculations performed in Section 4 show more quantitatively that the high-q objects stay far from Earth.   There are two tracks from the $\nu _6$ resonance to NEO space, one is a ``slow track'' involving encounters with Mars and another a ``fast track'' where the resonance kicks the eccentricity up on short time scales past the Mars Crosser region and into the Amors \citep{Morbidelli2002}. While we argue that it is possible that none of our high-MOID Q-types ever encountered Earth, it is only necessary that they have not encountered Earth in the past $\sim$1 My, the timescale at which space weathering occurs. If they have not crossed Earth or Venus recently, there are three main mechanisms currently identified that could be responsible for refreshing their surfaces: Mars encounters, YORP spin up, and collisions with other small bodies.

 %Mars: the 3 Q and 4 Q: objects
 \subsection{Mars Encounters}
\begin{figure}
  \centering
  \includegraphics[width=0.5\textwidth]{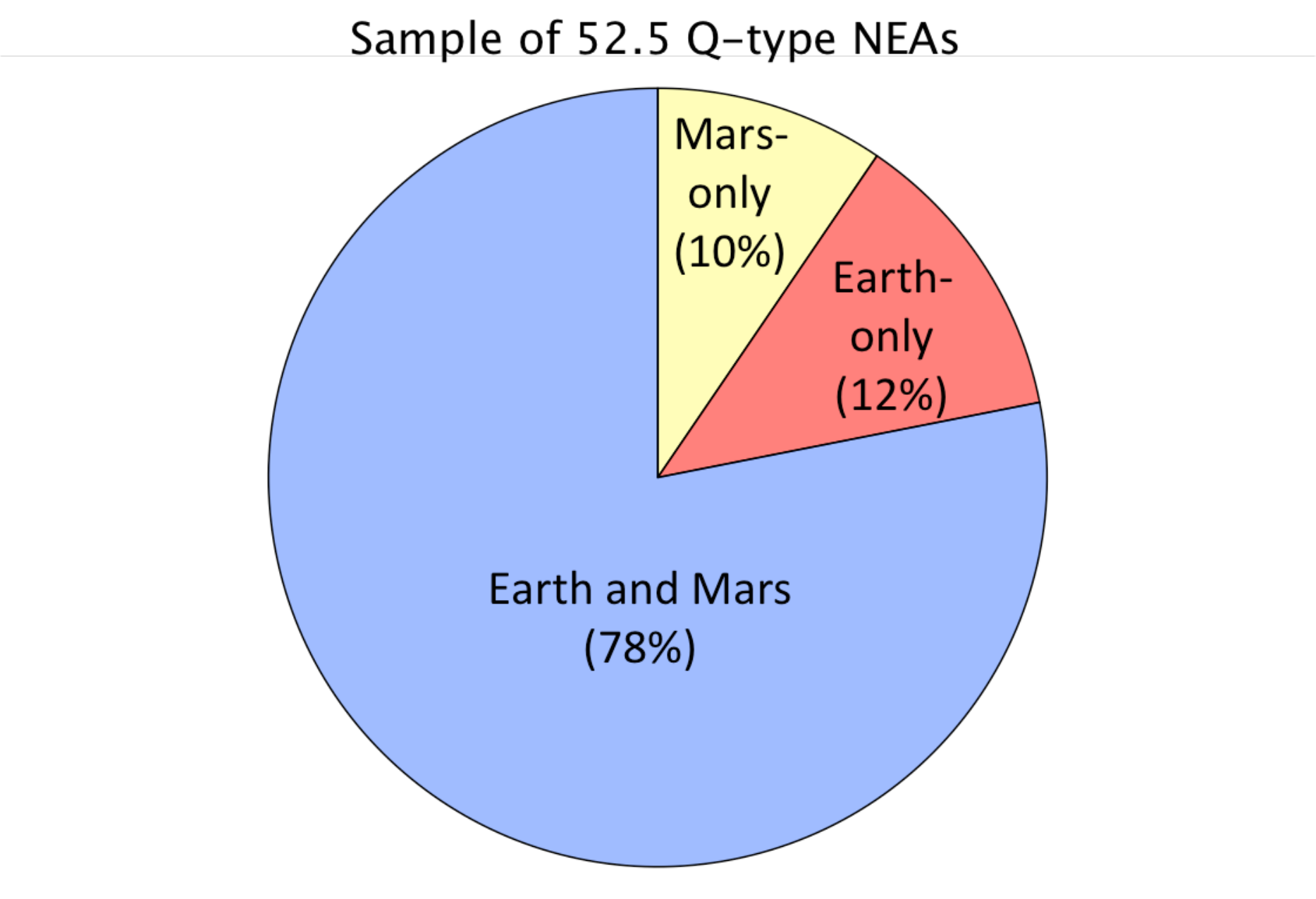}
  \caption[]{%
	Fraction of Q-type NEOs with orbits that encounter only Earth, only Mars, or both. There are no Q-types 
	in our sample that do not cross either planet. The majority of these bodies have orbits consistent with both
	Earth and Mars encounters making it challenging to distinguish between the two.} 
  \label{fig:pie}
\end{figure}

Of the subset of 7 Q-types (3 Q and 4 Q:) that do not intersect Earth, all have orbits that have the possibility to intersect Mars (we define a Mars intersection the same way for Earth: an object's MOID value with respect to Mars must be less than the Earth-Moon distance). The gravitational or tidal force that a body experiences at an equivalent distance from Earth or Mars is equal to the ratio of their masses. Thus a body would feel $\frac{1}{10}^{th}$ the force at a given distance from Mars compared to Earth. Conveniently, 10\% of our Q-type sample has orbits only allowing encounters with Mars. However, even though gravitational and tidal forces are weaker on Mars than Earth, asteroids encounter Mars more frequently increasing Mars' role. The fraction of all Q-types that encounter only Earth, only Mars, or both is depicted in Fig~\ref{fig:pie}. 

While 81\%  of our sample has orbits that are expected to have deep encounters with Earth, for Mars 94\% of our sample has the possibility of encounter. Of the S, Sq, and Q types in our sample, 15\% of objects that only encounter Mars are Q-type, 54\% that encounter Earth (and possibly Venus) but not Mars are Q-type, and of the objects that have the possibility to encounter both Mars and Earth 24\% are Q-type.

If we assume all 7 Q-types with a Mars-only crossing orbit have been refreshed due to Mars we can calculate the expected number of all our 64 Q-types due to Mars. This places an upper limit that as many as 58\% of the Q-types could be due to Mars. Most likely, however, the fraction of objects that have the possibility to encounter both Earth and Mars have a lower chance of being refreshed by Mars since some of those have orbits more favorable to Earth encounters.

Thus far, investigation of the properties of Q-type orbits has been restricted to the minimum distance a body might encounter a planet. The velocity of the encounter, however, may also play an important role in resurfacing since lower velocity encounters increase the likeliness of total disruption from tidal forces (Walsh, K., personal communication). While Q-types do not have to have experienced encounters deep enough for total disruption, the lower velocities may be more effective at surface refreshening as well.

%%%%-- BEGIN --%%%%%%%%%%--- Plot ---%%%%%%%%%%%%%%%%%%%%%%%%%%%%%%
%%%%--  END  --%%%%%%%%%%--- Plot ---%%%%%%%%%%%%%%%%%%%%%%%%%%%%%%

%YORP:
 \subsection{Collisions and YORP spin-up}
Based on the data in this work, all NEO Q-types can be explained by planetary encounter with Earth and Mars. Additional mechanisms such as YORP spin-up or collisions are possible processes that might be viable irrespective of Mars or Earth encounters, although we cannot disentangle them from planetary encounters. 
Here we review the possibility for YORP spin-up and collisions as mechanisms for surface refreshening.

YORP spin-up, which causes spin rates to increase and often for bodies to fission into two \citep{Bottke2006}, can be a potentially important process for refreshing surfaces since particles from the surface are moved toward the equator and in some cases, escape \citep{Walsh2008}. It is effective at larger heliocentric distances and is independent of planet-crossing orbits, as evident by the presence of asteroid pairs in the main belt \citep{Pravec2010}. For irregularly shaped bodies, YORP can cause the spin rate of an asteroid to either increase or decrease dramatically depending on the shape and spin orientation \citep{Vokrouhlicky2003}, and it is most effective on smaller bodies, such as the size range of $\sim$1km \citep{Bottke2006}, observed in the NEO population and including our high-MOID Amors.

While collisions between near-Earth asteroids have a relatively low probability, collisions with main-belt asteroids occur more frequently \citep{Bottke1994}. \citet{Bottke2005} calculate that a catastrophic disruption occurs with a 1 km asteroid every 1000 years, and for a 100 meter asteroid it occurs every day in the main belt.  The primary questions are: \textit{1. What size must the impactor be to refresh the surface of an asteroid of a given size?} and \textit{2. How frequently does such a collision occur?} It is only necessary to overturn surface grains to refreshen a surface, which involves much less energy than a larger collision that ejects material or causes global seismic activity.

To put into context an appropriate size for an impactor we calculate examples of the diameter of a projectile that would catastrophically disrupt (d$_{disrupt}$) target asteroids with diameters (D$_{target}$) of 100 m and 1 km. The equation from \citet{Bottke2005b} is:

\begin{equation}
d_{disrupt}=(\frac{2Q^*_D}{V^2_{imp}})^\frac{1}{3}D_{target}
\end{equation}

where $V^2_{imp}$ is the impact velocity which we will take to be 5 km/s from \citet{Bottke1994}, a conservative estimate since NEA velocities should be higher. $Q^*_D$ is the specific energy (erg/g) needed for catastrophic disruption, and the densities are assumed to be equal. $Q^*_D$ varies as a function of asteroid size (and other factors, such as density, porosity, and composition), and authors have estimated different values within roughly 2 orders of magnitude \citep[see Figures in ][]{Asphaug2002,Holsapple2002}. Using a range $Q^*_D$ determined from previous work we calculate projectile diameters for disruption: for a 100 m diameter target the impactor must have a diameter of 2 m or 4 m (for the case of $Q^*_D$ of 10$^6$ and 10$^7$, respectively) and for a 1km target d$_{disrupt}$ is 20 m or 40 m. These sizes are of course very rough estimates. For Eros, an impactor slightly larger than 1 km would be necessary for catastrophic disruption \citep{Melosh1997,Benz1999} and a minimum impactor diameter of $\sim$2 m would be necessary for a global seismic acceleration that exceeds the surface gravity \citep{Richardson2005}. While catastrophic disruption would certainly reset an asteroids' surface, it would take a much smaller impactor than calculated above to either gently flip grains over or to excavate or eject material. 

If collisions with such small bodies could provide enough energy to eject material the question then becomes: \textit{Why do we find any weathered surfaces at small sizes at all?} First, \citet{Marchi2012} proposed the concept of ``saturation'' whereby an object that experiences impact gardening (frequent small impacts that eject grains) is eventually covered by the reaccumulation of weathered particles. ``Sauteed'' is a descriptive term also representing the concept is that these grains have been overturned (or ``fried'') so many times that the surface is essentially entirely weathered down to a certain depth, and only a much stronger collision or other mechanism (such as YORP spin-up) can remove or overturn a sufficient amount of material to expose fresh grains. Second,  we \textit{assume} that the impacts actually cause material to be excavated. If a body is highly porous, as in the case for C-type asteroid (253) Mathilde for example, an impact could merely cause compaction rather than excavation \citep{Housen1999}. Although trends show that macroporosity for C-types is typically larger than for S-types \citep{Carry2012}, it is possible that compaction due to impact is relevant for the surfaces of some S-type asteroids. 

Another possibility for why we find weathered surfaces at all is related to the assumption that all S-complex asteroids are weathered. \citet{MotheDiniz2010} show that a number of ordinary chondrites are good spectral matches at visible wavelengths to some S-complex classes  \citep[particularly the Bus Sk and Sq classes,][]{Bus2002b}. In fact, they find that objects in these classes in the MB are on orbits with larger eccentricities than other S classes suggesting that they could be fresher due to higher probabilities of collisions \citep{MotheDiniz2010}. There is still much to understand about the compositional diversity and the extent and effects of space weathering on S-complex asteroids.

\citet{DellOro2011} perform an analysis of probability and velocity of collisions of NEOs with MBAs. They find a correlation between the degree of space weathering of an object (measured by spectral slope) and an NEO's probability and intensity of collision. Evidence of recent collisions in the main belt were seen from dust trails on asteroids (596) Scheila \citep{Bodewits2011,Jewitt2011b,Moreno2011,Ishiguro2011,Hsieh2012} and possibly P/2012 A$_2$ \citep{Jewitt2010,Snodgrass2010}. Other work, however, suggests that the dust trail of P/2012 A$_2$, because of its small size is more likely explained by the object reaching the spin rate limit due to YORP \citep{Marzari2011}.

Recent studies suggest that a combination of the YORP effect and collisions cause the current spin rate distribution of small main-belt asteroids. In fact, \citet{Marzari2011} suggest that it is the balance of the two effects that creates the population of very slow spin rates, they cannot be explained by one or the other alone. Additionally, preliminary work on the effect of YORP spin-up on the size distribution of the main belt at smaller sizes reveals that mass shedding through YORP is likely responsible for the faster erosion of small asteroids \citep{Rossi2012}. Their work shows that in the 1 km and smaller size regime, the fission rate due to YORP spin-up is comparable to the collision rate. A study from the Hayabusa mission suggests that asteroid Itokawa is losing tens of centimeters of surface material to space each year, an important amount of mass loss for the surface an NEO \citep{Nagao2011}.

YORP and collisions may collaborate to refresh asteroid surfaces. \citet{Scheeres2010}  find that for small asteroids, regolith may be stable even at high spin rates due to cohesive forces which dominate gravitational and other forces. They note, however, that a very small impactor, such as a micrometeorite could be sufficient to break the cohesive bonds of individual grains causing failure along stress fractures, disturbing the regolith.
 
% \textbf{Note: seismic shaking, eros, cohesion}

% thus likely small MB
It is likely that a combination of main-belt collisions and YORP spin-up are responsible for the fresh surfaces of a subset of our Q-types, although it is not required. This subset should be representative of the population of small (H$\gtrsim$16) main-belt asteroids which have similar collision probabilities and are possibly as susceptible to YORP spin-up (the NEOs receive more radiation than the main-belt objects, though the effect YORP will have on any object's spin rate is complex).  Small MBAs, however, are typically too faint to be observed spectroscopically.  Typical absolute magnitudes for MBAs observed spectroscopically are less than 13 (D$>$7km). The NEO potential Q-types with high perihelia have H magnitudes between H$\sim$ 16-18 (0.7km$<$D$<$2km). 

 %Review of christina work. short collisional lifetimes in MB
 \citet{Thomas2011} and \citet{Rivkin2011} performed a photometric survey of small Koronis family asteroids to search for space weathering trends. They find that for sizes smaller than 5 kilometers (H$\gtrsim$14), the average slope of these bodies decreases, suggesting that these smaller bodies have younger surfaces than the larger ones. \citet{Thomas2012} examined the average slope and band depth of asteroids within S-type families using data from the Sloan Digital Sky Survey. They again find average slope decreasing and band depth increasing with smaller sizes. \citet{MotheDiniz2008} spectroscopically detected a main-belt Q-type in a young family. A spectroscopic survey using a large telescope could potentially determine the fraction of Q-types among small MBAs which would allow us to better disambiguate among freshening mechanisms for the NEOs.

% why we don't see MB/MC Q
If NEOs experience Mars encounters or collisions with small MBAs that refresh their surfaces we would expect similar results for Mars Crossers (MCs, asteroids on orbits that cross Mars with perihelia greater than 1.3 AU) that spend a more substantial amount of their orbit in the main belt. We find no potential Q-types in our Mars Crosser sample, though there are only 47 S-complex MC objects compared with 249 S-complex NEAs. However, because 15\% of our non-Earth encountering NEOs were Q-type we could reasonably expect the same for MCs, in which case we would expect to find 7 Q-types in our sample of 47 objects. The lack of Q-types in our MC sample could be due to their larger sizes. By definition Mars Crossers have perihelia greater than 1.3 AU, limiting the size of objects we can observe from Earth. Most Mars crossers in the SMASS sample have perihelia between 1.4 and 1.7 AU and an H magnitude typically of 12-15.   We have only 4 S-complex MCs in our sample in the same size range as typical NEOs (H=16-19).

\section{Conclusion}

Ten percent of the Q-types in our sample have not experienced Earth encounter on recent timescales. Thus, the orbital distribution of Q-types suggests Earth encounter is not the only resurfacing mechanism that counteracts the effects of space weathering. These non-Earth encountering objects do cross the orbit of Mars and show low Mars-MOID values. We conclude that Mars is likely to play an important role in refreshing NEO surfaces due to its large mass and frequent asteroid encounters. With this dataset we cannot rule in YORP spin-up or main-belt collisions as additional mechanisms for surface refreshing nor can we constrain what fraction of NEOs could be refreshed by these two processes.  The effectiveness of these mechanisms could be further constrained by observations of small main belt asteroids.

\section*{Acknowledgments}
Thanks to Alessandro Morbidelli, Brett Gladman, Sarah Greenstreet, Bill Bottke, David Nesvorny, and Steve Slivan for useful discussions. We thank students Megan Mansfield and Michael Kotson for helping assemble the observation table. We thank Vishnu Reddy and Thais Mothe-Diniz for very helpful reviews that greatly improved the manuscript.
Observations reported here were obtained at the Infrared Telescope Facility, which is operated by the University of Hawaii under Cooperative Agreement NCC 5-538 with the National Aeronautics and Space Administration, Science Mission Directorate, Planetary Astronomy Program. 
This material is based upon work supported by the National Science Foundation under Grant 6920422. Any opinions, findings, and conclusions or recommendations expressed in this material are those of the authors and do not necessarily reflect the views of the National Science Foundation. 
All (or part) of the data utilized in this publication were obtained and made available by the The MIT-UH-IRTF Joint Campaign for NEO Reconnaissance. The IRTF is operated by the University of Hawaii under Cooperative Agreement no. NCC 5-538 with the National Aeronautics and Space Administration, Office of Space Science, Planetary Astronomy Program. The MIT component of this work is supported by NASA grant 09-NEOO009-0001, and previously by the National Science Foundation under Grant No. 0506716.

\clearpage

\section*{Appendix}

\begin{figure}
  \centering
  \includegraphics[width=\textwidth]{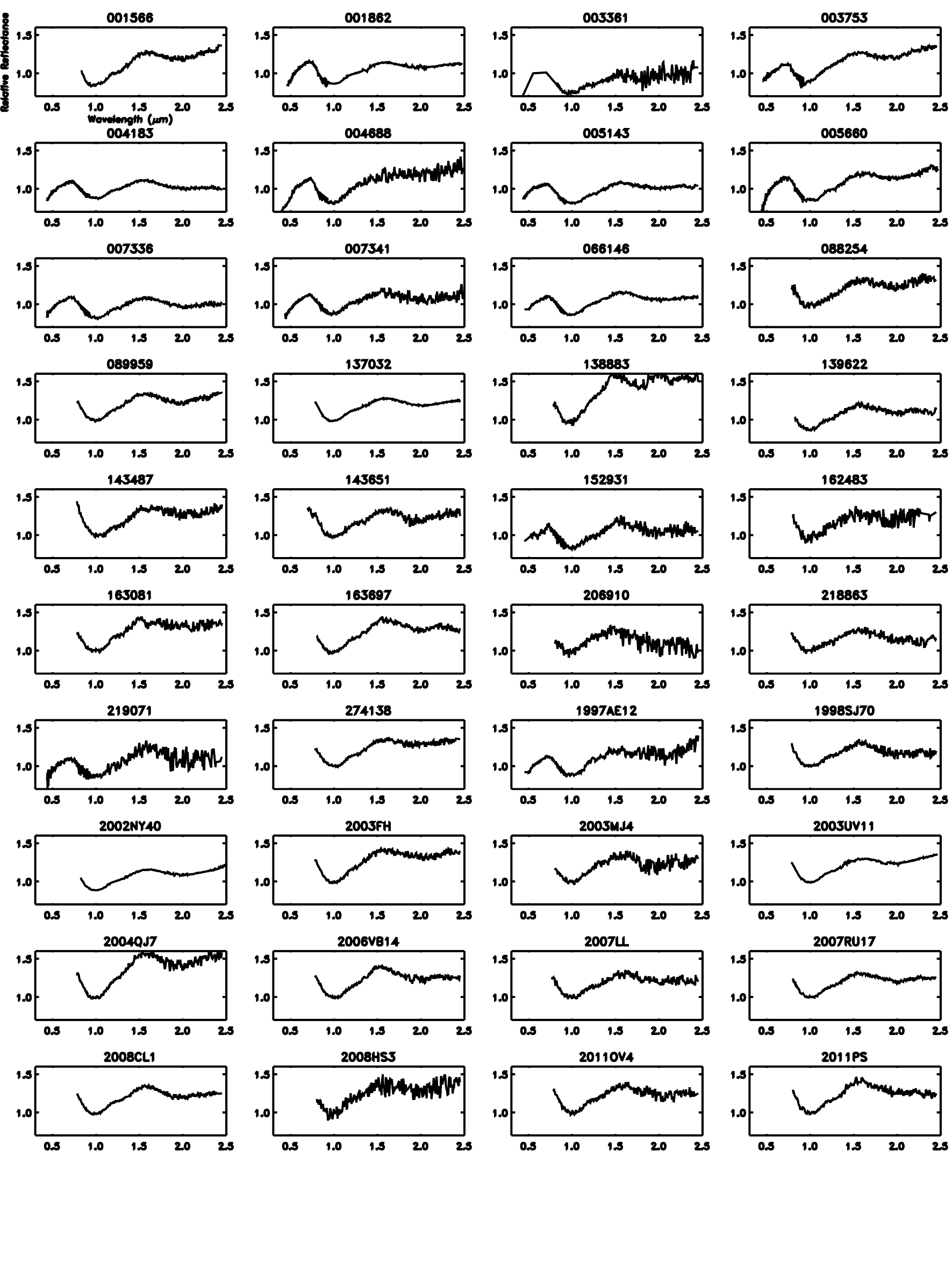}
  \caption[]{%
 Spectra of 41 Q-types NEOs in this work.} 
%  \label{fig: moid-mars}
\end{figure}

\clearpage

\begin{figure}
  \centering
  \includegraphics[width=\textwidth]{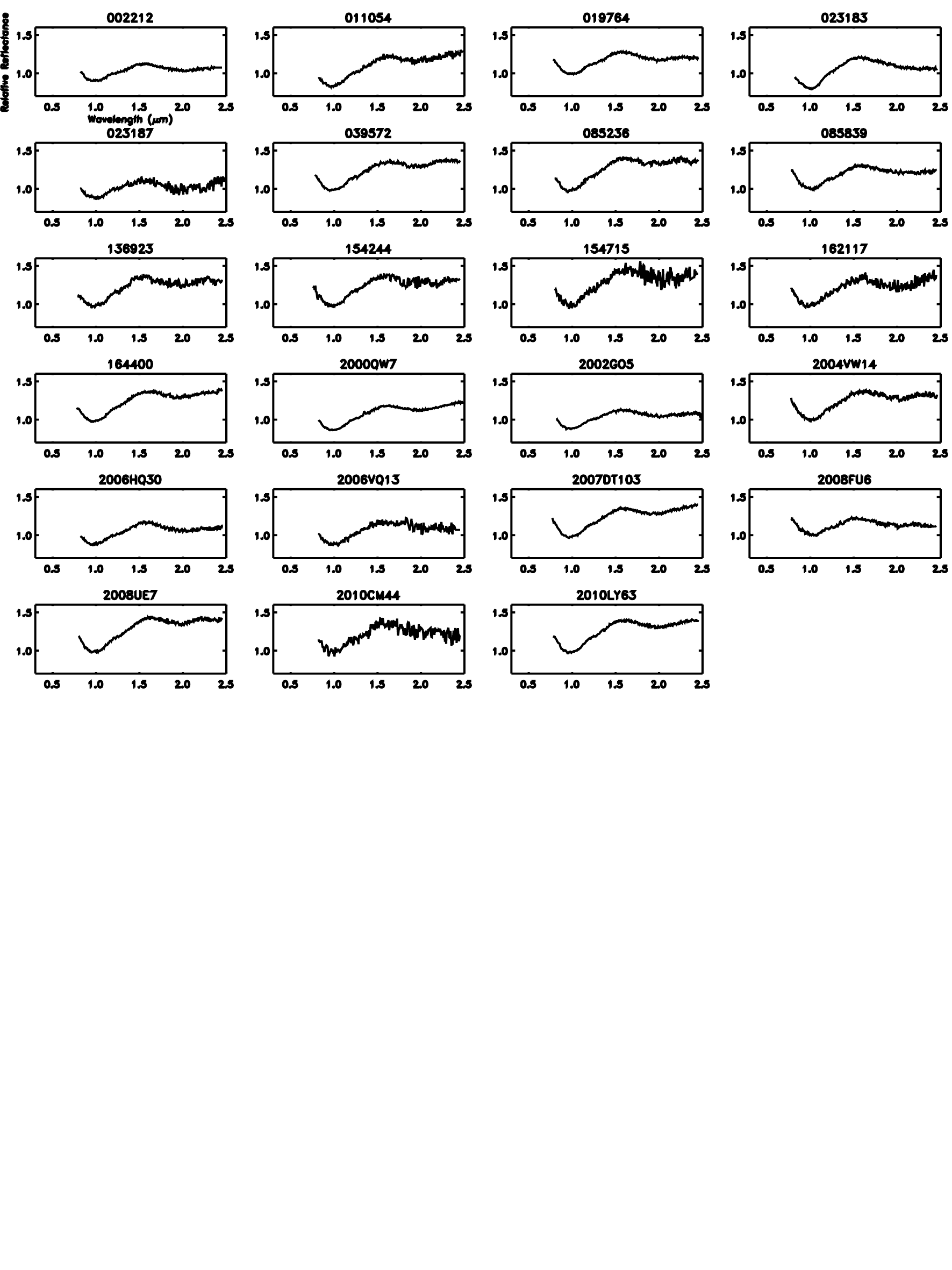}
  \caption[]{%
 Spectra of 23 Q: (Sq/Q-type) NEOs in this work.} 
%  \label{fig: moid-mars}
\end{figure}

\clearpage
\bibliographystyle{elsarticle-harv.bst}

\end{document}